# Coverage of highly-cited documents in Google Scholar, Web of Science, and Scopus: a multidisciplinary comparison

Alberto Martín-Martín[1] , Enrique Orduna-Malea[2] , Emilio Delgado López-Cózar[1]

## Abstract

This study explores the extent to which bibliometric indicators based on counts of highly-cited documents could be affected by the choice of data source. The initial hypothesis is that databases that rely on journal selection criteria for their document coverage may not necessarily provide an accurate representation of highly-cited documents across all subject areas, while inclusive databases, which give each document the chance to *stand on its own merits*, might be better suited to identify highly-cited documents. To test this hypothesis, an analysis of 2,515 highly-cited documents published in 2006 that Google Scholar displays in its *Classic Papers* product is carried out at the level of broad subject categories, checking whether these documents are also covered in Web of Science and Scopus, and whether the citation counts offered by the different sources are similar. The results show that a large fraction of highly-cited documents in the Social Sciences and Humanities (8.6%-28.2%) are invisible to Web of Science and Scopus. In the Natural, Life, and Health Sciences the proportion of missing highly-cited documents in Web of Science and Scopus is much lower. Furthermore, in all areas, Spearman correlation coefficients of citation counts in Google Scholar, as compared to Web of Science and Scopus citation counts, are remarkably strong (.83-.99). The main conclusion is that the data about highly-cited documents available in the inclusive database Google Scholar does indeed reveal significant coverage deficiencies in Web of Science and Scopus in several areas of research. Therefore, using these selective databases to compute bibliometric indicators based on counts of highly-cited documents might produce biased assessments in poorly covered areas.

## Acknowledgements

Alberto Martín-Martín enjoys a four-year doctoral fellowship (FPU2013/05863) granted by the Ministerio de Educación, Cultura, y Deportes (Spain).

Readers are still encouraged to send their feedback via e-mail (address of corresponding author is available in the footer), Twitter (including the link to the article, or the authors' handles: @albertomartin, @eomalea, @GScholarDigest), or PubPeer.

[1] Facultad de Comunicación y Documentación, Universidad de Granada, Granada, Spain.
[2] Universitat Politècnica de València, Valencia, Spain.
✉ Alberto Martín-Martín
   albertomartin@ugr.es

# Introduction

## The issue of database selection for calculating bibliometric indicators

It has been proposed that bibliometric indicators based on counts of highly-cited documents are a better option for evaluating researchers than using indicators such as the h-index (Bornmann & Marx, 2014; Leydesdorff, Bornmann, Mutz, & Opthof, 2011). A recent discussion held within the journal Scientometrics brought up this issue once again (Bornmann & Leydesdorff, 2018).

It is known that database selection affects the value that a bibliometric indicator takes for a given unit of analysis (Archambault, Vignola-Gagné, Côté, Larivière, & Gingrasb, 2006; Bar-Ilan, 2008; Frandsen & Nicolaisen, 2008; Meho & Yang, 2007; Mongeon & Paul-Hus, 2016). These differences are sometimes caused by diametrically opposed approaches to document indexing: indexing based on journal selection (Web of Science, Scopus), or inclusive indexing based on automated web crawling of individual academic documents (Google Scholar, Microsoft Academic, and other academic search engines). For an exhaustive commentary and bibliography on studies that compare the coverage and bibliometric indicators available in the previously mentioned databases (especially for studies that involve Google Scholar), we refer to Halevi, Moed & Bar-Ilan (2017), and Orduna-Malea, Ayllón, Martín-Martín, & Delgado López-Cózar (2015). Lastly, Delgado López-Cózar, Orduna-Malea, & Martín-Martín (forthcoming) presents a detailed summary of all studies published to date that discuss the differences between Google Scholar, Web of Science, and Scopus in terms of coverage and bibliometric indicators, and the correlations of citation-based indicators at various levels of aggregation[3].

Using databases in which document coverage depends on journal selection criteria (selective databases) to calculate indicators based on counts of highly-cited documents could produce biased assessments. This is because documents other than those published in journals selected by these databases could also become highly-cited. These documents could be books, reports, conference papers, articles published in non-selected journals... which could very well meet the same quality criteria as the documents covered in selective databases. Because it is not possible to predict which documents are going to become highly-cited before they are published, an inclusive database that gives each document the chance to *stand on its own merit* (Acharya, 2015), might in theory provide a better coverage of highly-cited documents than a selective database where document coverage is constricted to specific sources selected beforehand.

Compounded with the previous issue, there is the fact that Web of Science and Scopus, the most widely used selective databases for bibliometric analyses, are known to have poor coverage of areas in which research often has a local projection such as the Social Sciences and Humanities (Mongeon & Paul-Hus, 2016), as well as a bias against non-English publications (Chavarro, Ràfols, & Tang, 2018; van Leeuwen, Moed, Tijssen, Visser, & Van Raan, 2001). This goes against the principle of protecting "excellence in locally relevant research" in the Leiden Manifesto (Hicks, Wouters, Waltman, de Rijcke, & Rafols, 2015).

There is evidence to show that highly-cited documents are not only being published in elite journals. Acharya et al. (2014) found that, according to data from Google Scholar, the number of highly-cited documents published in non-elite journals had significantly grown between 1995 and 2013. They posited that this change was made possible by web search and relevance rankings, which meant that nowadays "finding and reading relevant articles in non-elite journals is about as easy as finding and reading articles in elite journals", whereas before web search, researchers were mostly limited to what they could browse in physical libraries, or to systems that only presented results in reverse chronological order. Martín-Martín, Orduna-Malea, Ayllón, and Delgado López-Cózar (2014) carried out an analysis of 64,000 highly-cited documents according to Google Scholar, published between 1950 and 2013. In this exploratory study they found that 49% of the highly-cited documents in the sample were not covered by the Web of Science. They also found that at least 18% of these 64,000 documents were books or book chapters (Martín-Martín, Orduna-Malea, Ayllón, & Delgado López-Cózar, 2016).

---

[3] Supplementary material to book chapter containing summary tables already available at: https://doi.org/10.17605/OSF.IO/PQR53

Google Scholar's *Classic Papers*

Since June 14<sup>th</sup> 2017, Google Scholar started providing a new service called *Classic papers*[4] which contains lists of highly-cited documents by discipline. Delgado López-Cózar, Martín-Martín, and Orduna-Malea (2017) explored the strengths and limitations of this new product.

The current version of Google Scholar's *Classic Papers* displays 8 broad subject categories. These broad categories contain, in total, 252 unique, more specific subject categories. Each specific subject category (from here on called subcategory) contains the top 10 most cited documents published in 2006. These documents meet three inclusion criteria: they presented original research, they were published in English, and by the time of data collection (May 2017, and therefore at least 10 years after their publication), they had at least 20 citations. Documents appear to have been categorized at the article level, judging by the fact that articles in multidisciplinary journals such as *Nature*, *Science*, or *PNAS* are categorized according to their respective topics. Appendix A provides a high-level comparison of how Google Scholar, Web of Science, and Scopus classify this sample of documents.

Despite the fact that, in line with Google Scholar's usual lack of transparency, there are many unanswered methodological questions about the product, like how the subject categorization at the document level was carried out, this dataset could shed some light on the differences in coverage of highly-cited documents in Google Scholar, Web of Science, and Scopus. The results may provide evidence of the advantages and disadvantages of selective databases and inclusive databases for the specific purpose of finding highly-cited documents.

## Research Questions

This study aims to answer the following research questions:
RQ1. How many highly-cited documents according to Google Scholar are not covered by Web of Science and Scopus? Are there significant differences at the level of subject categories?
RQ2. To the extent that coverage of highly-cited documents in these databases overlaps, are citation counts in Google Scholar similar in relative terms (rank orders) to those provided by Web of Science and Scopus?
RQ3. Which, out of Google Scholar, Web of Science, and Scopus, gives the most citations for highly-cited documents? Are there significant differences at the level of subject categories?

# Methods

In order to carry out the analysis, we first extracted all the information available in Google Scholar's *Classic Papers*. For this purpose, a custom script was developed which scraped all the relevant information, and saved it as a table in a spreadsheet file. The information extracted was:

- Broad subject categories and subcategories.
- Bibliographic information of the documents, including:
  - Title of the document, and URL pointing to the Google Scholar record for said document.
  - Authors (including URL to Google Scholar Citations profile when available), name of the publication venue, and year of publication.
  - Name and URL to Google Scholar Citations profile of showcased author (usually the first author, or the last author if the first doesn't have a public profile).
  - Number of citations the document had received when the product was developed (May 2017).

---

[4] https://scholar.googleblog.com/2017/06/classic-papers-articles-that-have-stood.html

A total of 2,515 records were extracted. All subcategories display the top 10 most cited documents, except the subcategory French Studies, in which only 5 documents were found with at least 20 citations.

Once the data from *Classic Papers* had been extracted, we proceeded to check how many of those 2,515 documents were also covered by Web of Science Core Collection, and Scopus. To do this, we used the metadata embedded in the URL that pointed to the Google Scholar record of the documents. In most cases, this URL contained the DOI of the document. Those DOIs were manually searched in the respective web interfaces of the other two databases, making sure that the documents that were found were actually the ones that were searched. In the cases when a DOI wasn't available in the URL provided by Google Scholar (only 105 records out of 2,515), and also when the DOI search wasn't successful, the search was conducted using the title of the document. If the document was found, its local ID in the database (the accession number in Web of Science, and the EID in Scopus), as well as its citation count was appended to the original table extracted from *Classic Papers*. For the documents that were not found, the cause why the document was not available was identified. The reasons identified were:

- The source (journal / conference) is not covered by the database.
- Incomplete coverage of the source (only some volumes or issues were indexed). A special case of this is when the source wasn't being indexed in 2006, but it started being indexed at a later date.
- The document has not been formally published: for the few cases (4) in which reports or preprints that were not eventually published made the list of highly-cited documents.

Data collection was carried out in June 2017, shortly after *Classic Papers* was launched. At the moment of writing this piece, searches in Web of Science and Scopus were carried out again to double-check that there had been no changes. It turned out that 2 additional documents were found in the Web of Science, and 7 additional documents were found in Scopus. These documents were not added to the sample, because by the time of the second search, they had had almost one additional year to accumulate citations and therefore comparisons of citation counts between sources would have not been fair.

Lastly, in order to clean the bibliographic information extracted from Google Scholar, which often presented incomplete journal or conference titles, we extracted the bibliographic information from CrossRef and DataCite using the available DOIs and content negotiation. For the cases when no DOI was available, the information was exported from Scopus, or added manually (mostly for the 79 documents which were not available in either of the databases).

To answer RQ1, the proportions of highly-cited documents in Google Scholar that were not covered in Web of Science and/or Scopus were calculated at the level of broad subject categories. Additionally, the most frequent causes why these documents were not covered are provided.

To answer RQ2, Spearman correlation coefficients of citation counts were calculated between the pairs of databases Google Scholar/Web of Science, and Google Scholar/Scopus. Correlation coefficients are considered useful in high-level exploratory analyses to check whether different indicators reflect the same underlying causes (Sud & Thelwall, 2014). In this case, however, the goal is to find out whether the same indicator, based on different data sources, provides similar relative values. Spearman correlations were used because it is well-known that the distributions of citation counts and other impact-related metrics are highly skewed (De Solla Price, 1976).

To answer RQ3, the average log-transformed citation counts for the three databases were calculated at the level of broad subject categories, and the normal distribution formula was used to calculate 95% confidence intervals for the log-transformed data (Thelwall, 2017; Thelwall & Fairclough, 2017).

The raw data, the R code used for the analysis, and the results of this analysis are openly available (Martín-Martín, Orduna-Malea, & Delgado López-Cózar, 2018).

# Results

## *RQ1. How many highly-cited documents according to Google Scholar are not covered by Web of Science and Scopus? What are the differences at the level of subject categories?*

Out of the 2,515 documents displayed in Google Scholar's *Classic Papers*, 208 (8.2%) were not covered in Web of Science, and 87 (3.4%) were not covered in Scopus. In total, 219 highly-cited documents were not covered either by Web of Science or Scopus. Among these, 175 of them were journal articles, 40 were conference papers, one was a report, and three were preprints. Regarding these preprints, all three are in the area of Mathematics. As far as we could determine, a heavily modified version of one of the preprints was published in a journal two years after the preprint was first made public, but the other two preprints have not been published in journals.

Significant differences in coverage were found across subject categories (Table 1). The areas where there are more highly-cited documents missing from Web of Science and Scopus are *Humanities, Literature & Arts* (28.2% in Web of Science, 17.1% in Scopus), and *Social Sciences* (17.5% in Web of Science, and 8.6% in Scopus). Moreover, Web of Science seems to be missing many highly-cited documents from *Engineering and Computer Science* (11.6%), and *Business, Economics & Management* (6.0%). The coverage of these last two areas in Scopus seems to be better (2.5% and 2.7% missing documents, respectively).

**Table 1. Number of highly-cited documents in Google Scholar that are not covered by Web of Science and/or Scopus, by broad subject areas**

| Subject category | N | Not in WoS | % | Not in Scopus | % |
|---|---|---|---|---|---|
| Humanities, Literature & Arts | 245 | 69 | 28.2 | 42 | 17.1 |
| Social Sciences | 510 | 89 (J: 88, R: 1) | 17.5 | 44 (J: 43, R: 1) | 8.6 |
| Engineering & Computer Science | 570 | 66 (J: 26, C: 40) | 11.6 | 14 (J: 10, C: 4) | 2.5 |
| Business, Economics & Management | 150 | 9 | 6.0 | 4 | 2.7 |
| Health & Medical Sciences | 680 | 19 | 2.8 | 2 | 0.3 |
| Physics & Mathematics | 230 | 5 (J: 2, P: 3) | 2.2 | 4 (J: 1, P: 3) | 1.7 |
| Life Sciences & Earth Sciences | 380 | 2 (J: 1, R: 1) | 0.5 | 2 (J: 1, R: 1) | 0.5 |
| Chemical & Material Sciences | 170 | 0 | 0 | 0 | 0 |

Unless otherwise specified, all missing publications are journal papers
J: journal paper; C: conference paper; P: preprint; R: report

Among the causes why some highly-cited documents were not covered in Web of Science and/or Scopus (Table 2), the most frequent one is that the journal or conference where the document was published was not covered in these databases in 2006, but it started been indexed at a later date (56% of the missing documents in Web of Science, and 49% of the missing documents in Scopus). Web of Science and Scopus do not practice backwards indexing except in special cases like the Emerging Sources Citation Index Backfile for documents published between 2005 and 2014, released on October 2017 and sold separately (Clarivate Analytics, 2017). Thus, documents published in journals before they are selected are missing from the databases.

**Table 2. Causes of highly-cited documents not being indexed in Web in Science and/or Scopus**

| The journal / conference where the document was published… | Web of Science (N = 208) | % | Scopus (N = 87) | % |
|---|---|---|---|---|
| … was not covered in 2006, but it was added at a later date (no backwards indexing) | 117 | 56 | 43 | 49 |
| … was being indexed in 2006, but coverage is incomplete (some volumes or issues are missing) | 50 | 24 | 12 | 14 |
| … is not covered by the database | 37 | 18 | 29 | 33 |
| The document is not formally published | 4 | 2 | 4 | 5 |

## RQ2. To the extent that coverage of highly-cited documents in these databases overlaps, are citation counts in Google Scholar similar in relative terms (rank orders) to those provided by Web of Science and Scopus?

If we focus exclusively in the documents that were covered both by Google Scholar and Web of Science, or by Google Scholar and Scopus, we find that the correlation coefficients are, in both cases, remarkably strong (Table 3).

**Table 3. Spearman correlation coefficients of citation counts between Google Scholar and Web of Science, and Google Scholar and Scopus, for highly-cited documents according to Google Scholar published in 2006, by broad subject categories**

| Subject category | GS-WoS | | GS-Scopus | |
|---|---|---|---|---|
| | N | Spearman corr. | N | Spearman corr. |
| Humanities, Literature & Arts | 176 | .84 | 203 | .89 |
| Social Sciences | 421 | .86 | 466 | .91 |
| Engineering & Computer Science | 504 | .83 | 556 | .92 |
| Business, Economics & Management | 141 | .89 | 146 | .92 |
| Health & Medical Sciences | 661 | .94 | 678 | .95 |
| Physics & Mathematics | 225 | .93 | 226 | .94 |
| Life Sciences & Earth Sciences | 378 | .97 | 378 | .98 |
| Chemical & Material Sciences | 170 | .99 | 170 | .99 |

confidence level: 95%
p-values < 0.0001

The weakest correlations of citation counts between Google Scholar and Web of Science are found in *Engineering & Computer Science* (.83), *Humanities, Literature & Arts* (.84), *Social Sciences* (.86), and *Business, Economics & Management* (.89), but even these are strong. Between Google Scholar and Scopus, correlations are even stronger than between Google Scholar and Web of Science in all cases. The weakest one is also found in the *Humanities, Literature & Arts* (.89). In the rest of the subject categories, the correlations are always above .90, reaching their highest value in *Chemical & Material Sciences* (.99).

## RQ3. Which, out of Google Scholar, Web of Science, and Scopus, gives the most citations for highly-cited documents?

Citation counts of highly-cited documents in Google Scholar are higher than citation counts in Web of Science and Scopus in all subject categories (Figure 1). Furthermore, the differences are statistically significant in all subject categories. They are larger in *Business, Economics & Management*, *Social Sciences*, and *Humanities, Literature & Arts*. The smallest difference that involves Google Scholar is found in *Chemical & Material Sciences*, where the lower bound of the 95% confidence interval for Google Scholar citation counts is closest to the higher bound of the confidence intervals for Scopus and Web of Science data.

**Figure 1. Average log-transformed citation counts of highly-cited documents according to Google Scholar published in 2006, based on data from Google Scholar, Web of Science, and Scopus, by broad subject categories**

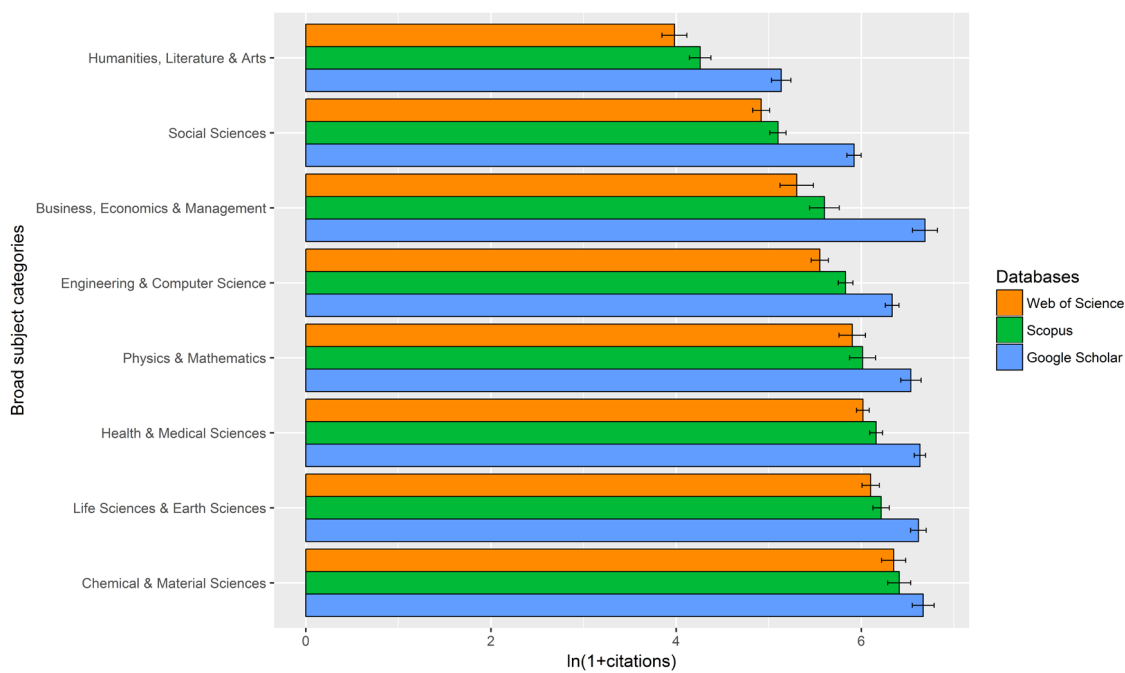

If we look at the differences between Web of Science and Scopus, we observe that, although the average of log-transformed citation counts is always higher in Scopus, the differences are statistically significant in only 4 out of 8 subject categories: *Engineering & Computer Science*, *Health & Medical Sciences*, *Humanities, Literature & Arts*, and *Social Sciences*. Even in these areas, the confidence intervals are very close to each other.

## Limitations

Google Scholar's *Classic Papers* dataset suffers from a number of limitations to study highly-cited documents (Delgado López-Cózar et al., 2017). An important limitation is the arbitrary decision to only display the top 10 most cited documents in each subcategory, when it is well-known that the number of documents published in any given year greatly varies across subcategories. Moreover, the dataset only includes documents written in English which presented original research, and published in 2006. Nevertheless, these 10 documents should be well within the limits of the top 10% most cited documents suggested by Bornmann and Marx (2014) to evaluate researchers, even in the subcategories with the smallest output. Further studies could analyze whether similar effects are also found for non-English documents, and documents published in years other than 2006.

For this reason, the set of documents used in this study can be considered as an extremely conservative sample of highly-cited documents. Thus, negative results in our analysis (no missing documents in Web of Science or Scopus), especially in subcategories with a large output, should not be considered conclusive evidence that these databases cover most of the highly-cited documents that exist out there. On the other hand, positive results (missing documents in Web of Science or Scopus) in this highly exclusive set should put into question the suitability of these databases to calculate indicators based on counts of highly-cited documents, especially in some areas.

Another limitation of this study is that, although it analyzes how many highly-cited documents in Google Scholar are not covered by Web of Science and Scopus, it does not carry out the opposite analysis: how many highly-cited documents in Web of Science and Scopus are not covered by Google Scholar. This analysis deserves its own separate study, but as a first approximation, we can consider the results of a recent working paper (Martín-Martín, Costas, van Leeuwen, & Delgado López-Cózar, 2018) in which a sample of 2.6 million documents covered by Web of Science where searched in Google Scholar. The study found that 97.6% of all articles and reviews in the sample were successfully found in Google Scholar. Also, it is worth noting that this study only searched documents in Google Scholar using their DOI, and made no further efforts to find documents that were not returned by this type of search. Therefore, it is reasonable to believe that most or all the documents covered by Web of Science are also covered by Google Scholar.

# Discussion and conclusions

The results of this study demonstrate that, even when only journal and conference articles published in English are considered, Web of Science and Scopus do not cover a significant amount of highly-cited documents in the areas of *Humanities, Literature & Arts* (28.2% in Web of Science, 17.1% in Scopus), and *Social Sciences* (17.5% in Web of Science, and 8.6% in Scopus). Additionally, a significant number of documents in *Engineering & Computer Science*, and *Business, Economics & Management* are also invisible to the Web of Science. In the case of Computer Science the cause is that Web of Science did not cover as many conference proceedings as Google Scholar and Scopus, even though this type of publication is an important part of the literature in this field. Therefore, bibliometric indicators based on counts of highly-cited documents that use data from these two databases may be missing a significant amount of relevant information.

Spearman correlation coefficients of citation counts based on Google Scholar and Web of Science, and Google Scholar and Scopus, for the 8 broad subject categories used in this study are remarkably strong: from .83 in *Business, Economics & Management* (GS-WoS), to .99 in *Chemical & Material Sciences* (both GS-WoS, and GS-Scopus). This evidence matches the results found in other studies (Delgado López-Cózar et al., forthcoming; Moed, Bar-Ilan, & Halevi, 2016), and is a step towards dispelling doubts about the possibility that documents that are highly-cited in Google Scholar but are not covered by Web of Science and/or Scopus are merely the product of unreliable citation counting mechanism in the search engine. Therefore, the notion that Google Scholar citation counts are unreliable at the macro level (Bornmann et al., 2009) does not seem to hold anymore. Although coverage of fields such as Chemistry in Google Scholar may have been poor in the past (Orduña-Malea, Martín-Martín, Ayllón, & Delgado López-Cózar, 2016; Vine, 2006), that issue seems to have been solved, as Harzing (2013) already reported, and as this study confirms.

Also, although it is well-known that Google Scholar contains errors, such as duplicate documents and citations, incomplete and incorrect bibliographic information (Delgado López-Cózar et al., forthcoming; Orduna-Malea, Martín-Martín, & Delgado López-Cózar, 2017), and that it is easy to game citation counts because document indexing is not subjected to quality control (Delgado López-Cózar, Robinson-García, & Torres-Salinas, 2014), these issues seem to have no bearing on the overall values of the citation counts of highly-cited documents. Further studies are needed to check whether these correlations hold for larger samples of documents. If that is the case, it would no longer be justified to dismiss Google Scholar's citation counts as unreliable on account of the bibliographic errors present in this source, at least in macro-level studies.

Lastly, Google Scholar is shown to provide significantly higher citation counts than Web of Science and Scopus in all 8 areas. *Business, Economics & Management*, *Humanities, Literature & Arts*, and *Social Sciences* are the areas where the differences are larger. Previous studies also pointed in this direction (García-Pérez, 2010; Levine-Clark & Gil, 2008; Meho & Yang, 2007; Mingers & Lipitakis, 2010). This indirectly points to the existence of a much larger document base in Google Scholar for these areas of research, and provides a reasonable explanation for the weaker Spearman correlation coefficients of citation counts in these areas. Further studies could focus on identifying the sources of the citing documents. Some studies have already analysed citing documents (sources, document types, languages, unique citations) in Google Scholar and compared them to the citations found by Web of Science and Scopus (Bar-Ilan, 2010; de Winter, Zadpoor, & Dodou, 2013; Kousha & Thelwall, 2008; Meho & Yang, 2007; Rahimi & Chandrakumar, 2014). These studies reported that after journal articles, a large proportion of the citations found only by Google Scholar came from conference papers, dissertations, books, and book chapters. However, these studies focused on specific case studies, and most of them were carried out more than five years ago. Therefore, an updated, in-depth, multi-discipline analysis of the sources of citations in Google Scholar (that examines aspects such as document types, languages, peer-review status…), as compared to other citation databases like Web of Science and Scopus is now warranted, and could further elucidate the suitability of each platform as sources of data for different kinds of bibliometric analyses.

All this evidence points to the conclusion that inclusive databases like Google Scholar do indeed have a better coverage of highly-cited documents in some areas of research than Web of Science (*Humanities, Literature & Arts*, *Social Sciences*, *Engineering & Computer Science*, and *Economics & Management*) and Scopus (*Humanities, Literature & Arts*, and *Social Sciences*). Therefore, using these selective databases to compute bibliometric indicators based on counts of highly-cited documents might produce biased assessments in those poorly covered areas. In the other areas (*Health & Medical Sciences*, *Physics & Mathematics*, *Life Sciences & Earth Sciences*, *Chemical & Material Sciences*) all three databases seem to have similar coverage and citation data, and therefore the selective or inclusive nature of the database in these areas does not seem to make a difference in the calculation of indicators based on counts of highly-cited documents.

Google Scholar seems to contain useful bibliographic and citation data in the areas where coverage of Web of Science and Scopus is deficient. However, although there is evidence that it is possible to use Google Scholar to identify highly-cited documents (Martin-Martin, Orduna-Malea, Harzing, & Delgado López-Cózar, 2017), there are other practical issues that may discourage the choice of this source: lack of detailed metadata (for example, author affiliations, funding acknowledgements are not provided), or difficulty to extract data caused by the lack of an API (Else, 2018). As is often the case, the choice of data source presents a trade-off (Harzing, 2016). The suitability of each database (selective or inclusive) therefore depends on the specific requirements of each bibliometric analysis, and it is important that researchers planning to carry out these analyses are aware of these issues before making their choices, because these assessments often have direct consequences on the careers of individual researchers (hiring, promotion, or funding decisions) or institutions (university rankings).

*Appendix A. Top 5 most common subject categories assigned by Web of Science and Scopus to highly-cited documents in Google Scholar, by Google Scholar broad subject categories*

| Google Scholar category: **Humanities, Literature & Arts** | | Google Scholar category: **Social Sciences** | |
|---|---|---|---|
| Web of Science categories (176 docs.) | Scopus categories (203 docs.) | Web of Science categories (421 docs.) | Scopus categories (466 docs.) |
| Area Studies (24) | Arts and Humanities (138) | Psychology (58) | Social Sciences (285) |
| Linguistics (21) | Social Sciences (127) | Education & Educational Research (57) | Arts and Humanities (97) |
| Psychology (18) | Psychology (17) | Business & Economics (56) | Medicine (76) |
| Literature (17) | Economics, Econometrics and Finance (11) | Government & Law (48) | Psychology (69) |
| Social Sciences – Other Topics (16) | Medicine (7) | Social Sciences – Other Topics (32) | Economics, Econometrics and Finance (49) |

| Google Scholar category: **Business, Economics & Management** | | Google Scholar category: **Engineering & Computer Science** | |
|---|---|---|---|
| Web of Science categories (141 docs.) | Scopus categories (146 docs.) | Web of Science categories (504 docs.) | Scopus categories (556 docs.) |
| Business & Economics (113) | Business, Management and Accounting (87) | Engineering (217) | Engineering (223) |
| Social Sciences – Other Topics (20) | Economics, Econometrics and Finance (70) | Computer Science (145) | Computer Science (158) |
| Public Administration (12) | Social Sciences (36) | Materials Science (56) | Materials Science (72) |
| Environmental Sciences & Ecology (9) | Arts and Humanities (12) | Chemistry (52) | Chemical Engineering (65) |
| Science & Technology – Other Topics (6) | Decision Sciences (10) | Science & Technology – Other Topics (44) | Social Sciences (61) |

| Google Scholar category: **Physics & Mathematics** | | Google Scholar category: **Health & Medical Sciences** | |
|---|---|---|---|
| Web of Science categories (225 docs.) | Scopus categories (226 docs.) | Web of Science categories (661 docs.) | Scopus categories (678 docs.) |
| Physics (74) | Physics and Astronomy (97) | General & Internal Medicine (170) | Medicine (482) |
| Mathematics (73) | Mathematics (89) | Science & Technology – Other Topics (80) | General (80) |
| Science & Technology – Other Topics (31) | General (31) | Surgery (53) | Biochemistry, Genetics and Molecular Biology (73) |
| Engineering (21) | Engineering (27) | Neurosciences & Neurology (36) | Social Sciences (32) |
| Mechanics (17) | Computer Science (25) | Psychology (24) | Nursing (32) |

| Google Scholar category: **Life Sciences & Earth Sciences** | | Google Scholar category: **Chemical & Material Sciences** | |
|---|---|---|---|
| Web of Science categories (378 docs.) | Scopus categories (378 docs.) | Web of Science categories (170 docs.) | Scopus categories (170 docs.) |
| Science & Technology – Other Topics (122) | Agricultural and Biological Sciences (122) | Chemistry (75) | Chemistry (85) |
| Environmental Sciences & Ecology (51) | General (118) | Science & Technology – Other Topics (34) | Biochemistry, Genetics and Molecular Biology (53) |
| Biochemistry & Molecular Biology (48) | Biochemistry, Genetics and Molecular Biology (89) | Materials Science (31) | Chemical Engineering (48) |
| Agriculture (37) | Environmental Science (61) | Biochemistry & Molecular Biology (19) | Materials Science (40) |
| Cell Biology (27) | Medicine (40) | Physics (18) | General (29) |